# The 3D behavior of a twisted flux tube expanding in the corona: reconnection, writhe and jets

Ehsan TAVABI, Serge KOUTCHMY and Ali AJABSHIRIZADEH

*Abstract* – We discuss some aspects of magnetic reconnection which could help in understanding many aspects of magnetic plasma interactions. We will show that the helical structure often observed in polar jets is a natural consequence of magnetic helicity conservation in 3D reconnection driven by a collision of two parts of an emerging flux tube within the single emerged loop. We perform 3D simulations by solving the time-dependent, ideal MHD equations with a uniform initial twist. We deduce that the emergence of highly twisted magnetic flux introduces several null points, which in turn causes reconnection between opposite directions of magnetic field within a single loop at nearest part and the release of the trapped twist in the form of a helical jet-like emission.

A wide range of dynamic phenomena in the solar atmosphere, as for example soft x-ray jets and specific flares etc, has been associated with magnetic reconnection occurring in a three dimensional magnetic null point topology [1].

There is a lot of evidence for magnetic reconnection in x-ray jets and it is also seen in the SXT (Soft X-ray Telescope) observations that about 10% of them exhibit helical structure [3 & 4]. Magnetic twist is a visible candidate for the initiation of jet [5 & 6]. Also there is some observational evidence in some jets, such as the lambda-shape, for a whip-like motion which is interpreted as a slingshot-like motion as a result of reconnection; even spinning motion can be interpreted as a result of reconnection between a twisted flux tube and an untwisted flux tube [2]. The clear demonstration of twisted structure from stereoscopic observations of polar jets compares favorably with synthetic images from a recent MHD simulation invoking magnetic untwisting as their driving mechanism reported by [5].

In order to understand the dynamical evolution of an emerging flux ropes entering into an ambient atmosphere, we integrate the low-β compressible ideal MHD equations solved using a fourth-order Runge-Kutta method. We solved the equations and assume an ideal gas law, no radiative cooling, and no heat conduction. A Cartesian grid of size 8×8 Mm$^2$ in the horizontal directions and 12 Mm in the vertical Z direction, was chosen and equations resolved in 64×64×100 grid points.

The flux ropes are uniformly twisted along its cross-section, the twist is chosen right–handed, with the field lines winding for about 5 turns.

We change all variable scales to dimensionless variables by choosing against initial photospheric value for density, velocity, pressure and its scale (about of 170 km) and also the equilibrium field components having the form

$$B_r = B_0 J_1(\Lambda_B r) e^{-\Lambda_B z},$$
$$B_\varphi = \alpha B_0 J_1(\Lambda_B r) e^{-\Lambda_B z},$$
$$B_z = B_0 J_0(\Lambda_B r) e^{-\Lambda_B z}.$$

$B_0$ is the magnetic field at Z=0, the magnetic scale height is $\Lambda_B$ and J is the zero and first order of Bessel function. The sign of the azimuthal component of the magnetic field determines if the tube has left or right-handed twist and is shown by α.

The simulation shows the formation of a magnetic null point above the rising flux tubes and the subsequent breaking of the field with strong twist developing in one portion of the loop. The other portion retracts rapidly back toward the surface.

Manuscript received 30 July 2010; revised 18 March 2011.
E. Tavabi is with physics Department, Payame Noor University, 19395-4697 Tehran, I, R. of Iran, S. Koutchmy is with Institut d'Astrophysique de Paris (CNRS) & UPMC, UMR 7095, 98 Bis Boulevard Arago, F-75014 Paris, France, E. Tavabi and Ali Ajabshirizadeh are Research Institute for Astronomy & Astrophysics of Maragha (RIAAM), 55134-441 Maragha, Iran
Work was supported by RIAAM.
Publisher Identifier S XXXX-XXXXXXX-X

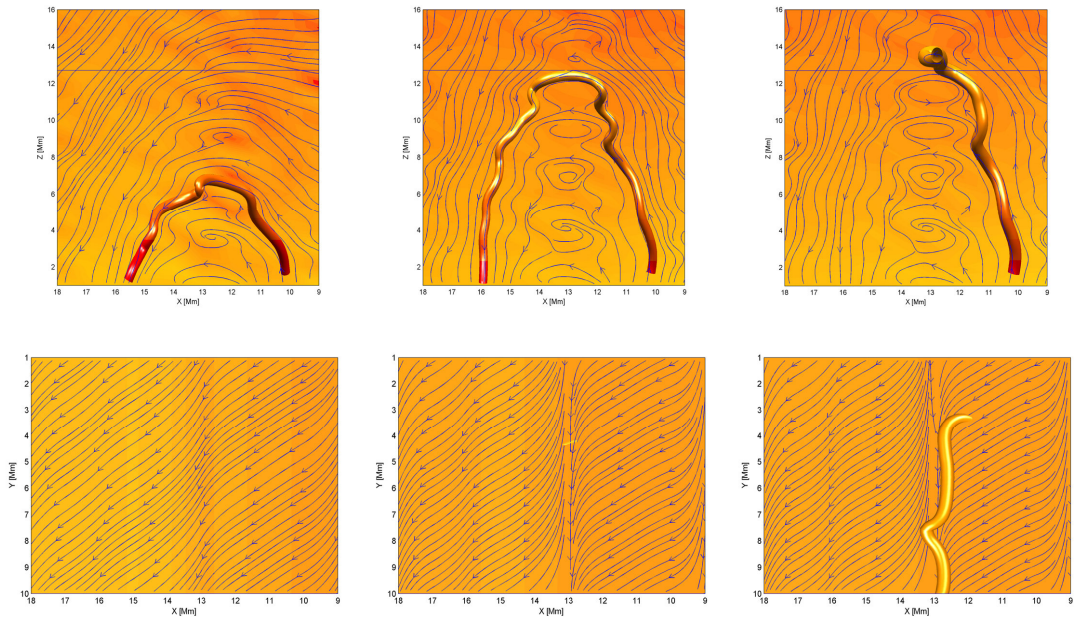

Fig. 1. Top panels: from left to right, evolution of the rope. Arrows on lines mark the magnetic field lines around and inside the loop. The bottom panels show a cross section in the horizontal direction of the top panels as shown by the line near z= 13 Mm. In these panels (left to right) we illustrate the flux tube and its subsequent rise towards the null points and then one step in the reconnection process.

In conclusion, we addressed the formation of 3D null point topologies during the expansion of twisted magnetic ropes, by means of reconnection between oppositely directed magnetic field lines along the single null-point, sometimes called a singular point of the "focus" or a "rosette" type, which corresponds to a singular point of the tangential field of the node with an added azimuthal field. The basic theory for using helicity conservation to determine the evolution of magneto-plasmas has been developed by Taylor. According to the Taylor theory [7], the helicity is the only topological quantity that is nearly conserved during reconnection. Our numerical simulations agree well with this statement. Our figure shows the appearance of several nullpoints along the Z axis (arrowheads mark the direction of the magnetic field, which are assumed to carry surge material), but the reconnection site is only produced when the upward and the downward field lines are located sufficiently nearby to each other, so apart from the assumed twist profile, the expansion of the rope is very important.

## REFERENCES

[1] E. R. Priest & T. G. Forbes, "Magnetic Reconnection: MHD Theory and Applications", (New York: Cambridge Univ. Press), (2000)

[2] R. C. Canfield, et al. "Hα Surges and X-ray Jets in AR 7260", ApJ, **464**, 1016, (1996)

[3] Shibata K, Ishido Y, Acton L *et al* 1992 *Publ. Astron. Soc. Japan* **44,** L173

[4] Shimojo M *et al Publ. Astron. Soc. Japan* **48,** 123, (1996)

[5] S. Patsourakos, et al., "*STEREO* SECCHI STEREOSCOPIC OBSERVATIONS CONSTRAINING THE INITIATION OF POLAR CORONAL JETS", ApJ, **680**, L73, (2008)

[6] Filippov, B., Golub, L., Koutchmy, S., 2009, Solar Phys., **254**, 259

[7] Taylor, J. B., Relaxation and magnetic reconnection in plasmas, in Rev. Mod. Phys., **58**, 741, (1986).